\documentclass[conference]{IEEEtran}

\usepackage[top=0.7in, left=0.68in, bottom=1.01in, right=0.68in]{geometry}

\IEEEoverridecommandlockouts
\usepackage{cite}
\usepackage{amsmath,amssymb,amsfonts}
\usepackage{algorithmic}
\usepackage{graphicx}
\usepackage{textcomp}
\usepackage[table]{xcolor}
\usepackage{cite}
\usepackage{balance}
\usepackage{bm}
\usepackage{algorithm}
\usepackage{algorithmic}
\usepackage{subcaption}

\begin{document}
	
	\bstctlcite{IEEEexample:BSTcontrol}
	
	\title{Deep Learning based Multi-User Power Allocation and Hybrid Precoding in Massive MIMO Systems
		\thanks{
			{This work was partially supported by InterDigital Inc. and the Natural Sciences and Engineering Research Council of Canada (NSERC).}
		}
	}
	
	\author{\IEEEauthorblockN{Asil Koc, Mike Wang, Tho Le-Ngoc}
		\IEEEauthorblockA{
			Department of Electrical and Computer Engineering, McGill University, Montreal, QC, Canada \\
			Email: asil.koc@mail.mcgill.ca,
			siyu.wang5@mail.mcgill.ca,
			tho.le-ngoc@mcgill.ca
			\vspace{-3ex}
		}
	}
	
	\maketitle
	
	\begin{abstract}
		
	This paper proposes a deep learning based power allocation (DL-PA) and hybrid precoding technique for multi-user massive multiple-input multiple-output (MU-{m}MIMO) systems. 
		We first utilize an angular-based hybrid precoding technique for reducing the number of RF chains and channel estimation overhead.
	{Then}, we develop the DL-PA algorithm via a fully-connected deep neural network (DNN).
		 DL-PA has two phases: (i) offline supervised learning with the optimal allocated powers obtained by particle swarm optimization based PA (PSO-PA) algorithm, (ii) online power prediction by the trained DNN.
		In comparison to the computationally expensive PSO-PA, it is shown that DL-PA greatly reduces the runtime by $98.6\%$-$99.9\%$, while closely achieving the optimal sum-rate capacity.
	It makes DL-PA a promising algorithm for the real-time online applications in MU-{m}MIMO systems.

	\end{abstract}
	

	\section{Introduction}
	
	\IEEEPARstart{M}{illimeter} wave (mmWave) has been considered as a promising candidate for the fifth-generation (5G) and beyond {for its large available} bandwidth \cite{Uwaechia2020}.
		Also, {its} shorter wavelengths are appealing for massive multiple-input multiple-output ({m}MIMO) technology
	{since it} enables the implementation of large antenna arrays in relatively smaller physical dimensions \cite{5G_Mas_MIMO_3}.
		{On the other hand}, {m}MIMO technology
		alleviates the severe path loss effect in mmWave communications via high beamforming gain.
	
	For multi-user downlink transmission, the conventional MIMO systems generally consider the single-stage fully-digital precoding (FDP) \cite{Mass_MIMO_Precoding_Survey}. However, FDP causes two major challenges for multi-user {m}MIMO (MU-{m}MIMO) systems:
		(i) the high hardware cost/complexity with the requirement of one dedicated power-hungry radio frequency (RF) chain per each antenna,
		(ii) large channel estimation overhead size \cite{Mass_MIMO_Hybrid_Survey}.
	{Alternatively}, two-stage hybrid precoding (HP) interconnects the digital baseband(BB)-stage and analog RF-stage with significantly reduced number of RF chains\cite{Mass_MIMO_Hyb_Survey,ASIL_FD_SIC,9507556}.
	Also, an angular-based HP (AB-HP) technique is developed in \cite{ASIL_ABHP_Access}, where analog RF-stage via is designed the slow time-varying angle-of-departure (AoD) information. Thus, AB-HP addresses both aforementioned challenges by decreasing the channel estimation overhead and the number of RF chains.
		On the other hand, multi-user power allocation (PA) is a non-convex optimization problem due to the effect of inter-user interference \cite{Emil_PA_NonConvex}. 
	Recently, \cite{ASIL_PSO_PA_WCNC} proposes an iterative particle swarm intelligence based PA (PSO-PA) algorithm for maximizing the overall system capacity in MU-{m}MIMO systems.
		Although it is shown that PSO-PA achieves the globally optimal system capacity, it requires longer runtime as the optimization space (i.e., number of users) increases.
	
	As a key driving force for artificial intelligence (AI),
	deep learning has been successfully applied in many
	fields including computer vision, speech recognition and natural language processing \cite{lecun2015deep}.
		Hence, the success of deep learning also motivates its applications in wireless communication systems \cite{DL_Survey_Wireless_2,DL_Survey_Wireless,ASIL_Xiaoyi_DL_CE}.
	For instance, deep learning has been applied for signal detection \cite{DL_Survey_Wireless_2}, resource management \cite{DL_Survey_Wireless}, channel estimation \cite{ASIL_Xiaoyi_DL_CE}. 
		Our ultimate goal is to investigate deep learning for a low-complexity PA technique achieving near-optimal system capacity with acceptable runtime considering real-time applications in MU-{m}MIMO systems with HP.	
	
	In this paper, we propose a novel low-complexity deep learning based PA (DL-PA) algorithm in MU-{m}MIMO systems utilizing HP architecture.
		We first employ AB-HP for the downlink transmission to reduce the number of RF chains and the channel estimation overhead size. 		
	Then, the proposed DL-PA is built via a fully-connected deep neural network (DNN).
		There are two phases in DL-PA: (i) offline supervised learning via the optimal allocated powers calculated with PSO-PA, (ii) online power prediction via the trained DNN.
		Numerical results present that DL-PA nearly achieves the optimal sum-rate capacity calculated by PSO-PA (e.g., $96.5\%$-$99.7\%$ of optimal capacity).
	Also, the runtime of PSO-PA is remarkably reduced by $98.6\%$-$99.9\%$ via DL-PA, which is essential regarding the real-time online applications.

	The rest of this paper is organized as follows. 
		Section \ref{SEC_SYS} expresses the system model. 
	Section \ref{SEC_HP} introduces AB-HP.
			Section \ref{SEC_DL_PA} presents the proposed DL-PA.
	After the illustrative results in Section \ref{SEC_RESULTS}, the paper is concluded in Section \ref{SEC_CONC}.

	\section{System Model}\label{SEC_SYS}
	
\begin{figure}[!t]
		\centering
		\includegraphics[width = \columnwidth]{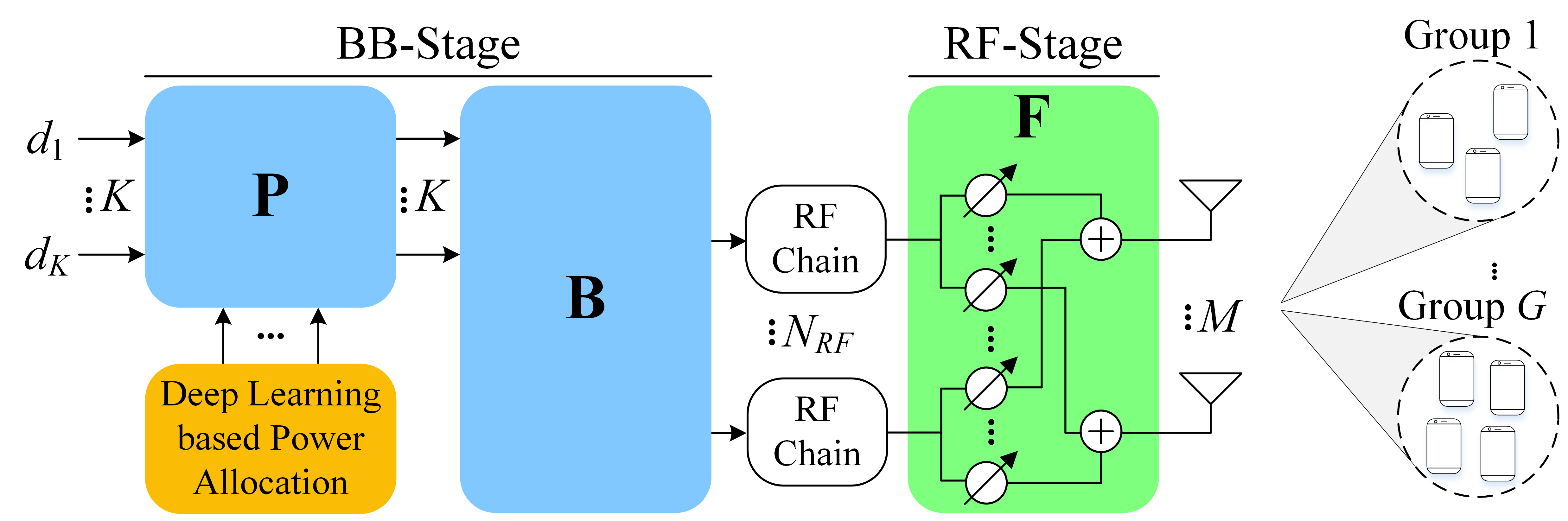}
		\caption{Massive MIMO system with hybrid precoding.}
		\label{fig1_System}
	\end{figure}

	A single-cell MU-{m}MIMO system is modeled for the downlink transmission as illustrated in Fig. \ref{fig1_System}. Here, the base station (BS) is equipped with a uniform rectangular array (URA) having $M=M_x\times M_y$ antennas\footnote{
		In the URA structure, $M_x$ and $M_y$ are the number of antennas along $x$-axis and $y$-axis, respectively. 
			Different from the widely considered uniform linear array (ULA), URA (i) fits a larger number of antennas in a two-dimensional (2D) grid, (ii) enables three-dimensional (3D) beamforming \cite{ASIL_ABHP_Access}. 
	}
	to serve $K$ single-antenna user equipments (UEs) clustered in $G$ groups. 
	
	As presented in Fig. \ref{fig1_System}, the RF-stage and BB-stage are interconnected via $N_{RF}$ RF chains to reduce the hardware cost/complexity (i.e., $K\le N_{RF}\ll M$).
	First, the analog RF beamformer ${\bf F}\in\mathbb{C}^{M\times N_{RF}}$ is developed via the low-cost phase-shifters for the RF-stage.
	Second, the digital BB precoder
	${\bf B}=\left[{\bf b}_1,\cdots,{\bf b}_K \right]\in\mathbb{C}^{N_{RF}\times K}$
	and 
	the multi-user PA matrix 
	${\bf P} =\textrm{diag}\left(\sqrt{p_1},\cdots,\sqrt{p_K}\right)\in\mathbb{R}^{K\times K}$ are constructed for the BB-stage, where ${\bf b}_k\in\mathbb{C}^{N_{RF}}$ and $p_k$ are the BB precoder vector and the non-negative allocated power for the $k^{th}$ UE, respectively.
	Hence, 
	the transmitted downlink vector is defined as 
	$
	{\bf s} = {\bf FBPd} \in \mathbb{C}^{M}
	$,
	where
	${\bf d} = \left[d_1,\cdots,d_K\right]\in\mathbb{C}^{K}$ is the data signal vector with $\mathbb{E}\big\{{\bf dd}^H\big\}= {\bf I}_K$.
	It is important to mention that ${\bf s}\in \mathbb{C}^{M}$ satisfies the total transmit power constraint of $P_T$ (i.e., $\mathbb{E}\big\{\big\|{\bf s}\big\|^2_2\big\}\le P_T$).

	According to the 3D geometry-based mmWave channel model \cite{Uwaechia2020} and the URA structure \cite{ASIL_ABHP_Access}, the channel vector for the $k^{th}$ UE is defined as follows:
	\begin{equation}\label{eq_h_k}
		{\bf{h}}_k^T
		=
		\sum_{l=1}^{Q} \tau^{-\eta}_{k_l}z_{k_l}{\bm{\phi }}^T 
		\big(
		{{\gamma _{x,k_l}},{\gamma _{y,k_l}}}
		\big) 
		= {\bf{z}}^T_k{{\bf{\Phi }}}_k
		\in
		\mathbb{C}^M,
	\end{equation}
	where 
	$Q$ is the number of paths,
	$\tau_{k_l}$ and $z_{k_l}\sim\mathcal{CN}\big(0,{1}/{Q}\big)$ are respectively the distance and complex path gain of $l^{th}$ path,
	$\eta$ is the path loss exponent,
	${\bm{\phi }}\big( \cdot,\cdot\big)\in \mathbb{C}^M$ is the phase response vector,
	$\gamma_{x,k_l}=\sin\left(\theta_{k_l}\right)\cos\left(\psi_{k_l}\right)$
	and
	$\gamma_{y,k_l}=\sin\left(\theta_{k_l}\right)\sin\left(\psi_{k_l}\right)$
	are the coefficients reflecting the elevation AoD (EAoD) and azimuth AoD (AAoD) for the corresponding path.
	Here,
	$\theta_{k_l}\in\big[\theta_k - \delta_k^\theta, \theta_k + \delta_k^\theta\big]$ is the EAoD with mean $\theta_k$ and spread $\delta_k^\theta$, 
	$\psi_{k_l}\in\big[\psi_k - \delta_k^\psi, \psi_k + \delta_k^\psi\big]$ is the AAoD with mean $\psi_k$ and spread $\delta_k^\psi$.
	Then, the phase response vector is modeled as \cite{ASIL_ABHP_Access}:
	\begin{equation}\label{eq_phase_vector}
		\begin{aligned}
			{\bm{\phi}}\hspace{-0.5ex}\left( {{\gamma_x, \gamma_y}} \right) \hspace{-0.5ex}&=\hspace{-0.75ex} \big[ {1,{e^{ -j2\pi d  {{\gamma_x }} }}, \cdots,{e^{ -j2\pi d\left( {{M_{x}} - 1} \right) {{\gamma_x}} }}} \big]^T\\
			&\otimes \hspace{-0.5ex}\big[ {1,{e^{-j2\pi d  {{\gamma_y }} }}, \cdots,{e^{-j2\pi d\left( {{M_{y}} - 1} \right) {{\gamma_y}} }}} \big]^T\in \mathbb{C}^M,
		\end{aligned}
	\end{equation}
	where $d$ is the antenna spacing normalized by wavelength.
	The instantaneous channel vector expressed in \eqref{eq_h_k} is a function of the fast time-varying path gain vector 
	${\bf z}_k = \big[\tau^{-\eta}_{k_1}z_{k_1},\cdots,\tau^{-\eta}_{k_Q}z_{k_Q}\big]^T\in\mathbb{C}^Q$
	and
	slow time-varying phase response matrix 
	${{\bf{\Phi }}}_k\in\mathbb{C}^{Q\times M}$ based on AoD information.
	
	Afterwards, the received signal at the $k^{th}$ UE is written as:
	\begin{equation}
		\begin{aligned}
			r_k
			&
			=
			{\bf h}_k^T {\bf s} + n_k
			=
			{\bf h}_k^T {\bf FBPd} + n_k\\
			&
			=
			\underbrace{\sqrt{p_k}{\bf h}_k^T {\bf Fb}_kd_k}_{\textrm{Desired Signal}}
			+
			\underbrace{\sum\nolimits_{t\ne k}^{K}\sqrt{p_t}{\bf h}_k^T {\bf Fb}_t d_t}_{\textrm{Inter UE Interference}}
			+ 
			n_k,
		\end{aligned}
	\end{equation}
	where
	$n_{k}\sim\mathcal{CN}\big(0,\sigma_n^2\big)$ is the  circularly symmetric complex Gaussian noise. After some mathematical manipulations, we derive the instantaneous signal-to-interference-plus-noise-ratio (SINR) at the $k^{th}$ UE as follows:
	\begin{equation}\label{eq_SINR}
		\begin{aligned}
			\textrm{SINR}_{k}
			\hspace{-0.35ex}
			\left(
			{\bf F},
			{\bf B},
			{\bf P}\right)
			=
			\frac
			{p_k\left|{\bf h}_k^T{\bf Fb}_k\right|^2}
			{\sum_{t\ne k}^{K}p_t\left|{\bf h}_k^T{\bf Fb}_t\right|^2 
				+
				\sigma_n^2}.
		\end{aligned}
	\end{equation}
	Then, the ergodic sum-rate capacity is calculated as 
	$R_{\textrm{sum}} = \mathbb{E}\big\{\sum_{k=1}^{K}
	\log_2
	\big[1+ 
	\textrm{SINR}_{k}
	\hspace{-0.35ex}
	\left(
	{\bf F},
	{\bf B},
	{\bf P}\right)
	\big]\big\}$.
	For maximizing the system capacity, we formulate the optimization problem as:
	\begin{equation}\label{eq_OPT_1}
		\begin{aligned}
			\max_{
				{\bf F},
				{\bf B},
				{\bf P}
			} ~& 
			\sum_{k=1}^{K}
			\log_2
			\left(
			1
			+
			\frac
			{p_k\left|{\bf h}_k^T{\bf Fb}_k\right|^2}
			{\sum_{t\ne k}^{K}p_t\left|{\bf h}_k^T{\bf Fb}_t\right|^2 
				+
				\sigma_n^2}
			\right)\\
			\textrm{s.t.}~~~
			&
			C_1: \mathbb{E}\big\{\hspace{-0.5ex}\left\|{\bf{s}}\right\|^2_2\hspace{-0.5ex}\big\} = \sum_{k = 1}^K {{p_{k}}{\bf{b}}_{k}^H{{\bf{F}}^H}{\bf{F}}{{\bf{b}}_{k}}} \le {P_T},\\
			&
			C_2: p_{k}\ge 0, \forall k,\\
			&
			C_3: \big|\left[{\bf{F}}\right]_{i,j}\big|=\frac{1}{\sqrt{M}},\forall i,j,\\
		\end{aligned}
	\end{equation}
	where
	$C_1$ and $C_2$ indicate the total and per UE transmit power constraints, respectively, $C_3$ refers {to} the constant modulus (CM) constraint due to the utilization of phase-shifters at the RF-stage.
	However, it is a non-convex optimization because of two reasons: (i) the allocated powers entangled with each other \cite{Emil_PA_NonConvex},  (ii) the CM constraint at the analog RF beamformer \cite{Mass_MIMO_Hyb_Survey}. 
	Thus, we sequentially design the hybrid precoding architecture illustrated in Fig. \ref{fig1_System}. First, the analog RF beamformer and the digital BB precoder are designed based on AB-HP technique in Section \ref{SEC_HP}, then the multi-user PA matrix is developed via the proposed deep learning based PA (DL-PA) algorithm in Section \ref{SEC_DL_PA}.

	\section{Angular-Based Hybrid Precoding (AB-HP)}\label{SEC_HP}
	
	Throughout this section, our ultimate goals are to
	(i) reduce the number of RF chains, 
	(ii) decrease the channel estimation overhead,
	(iii) mitigate the inter UE interference 
	via AB-HP technique for MU-{m}MIMO systems.
	
	
	\subsection{Analog RF Beamformer}\label{SSEC_RF}
	
	We construct the analog RF beamformer by focusing the signal energy in the desired direction via the slow-time varying AoD information.
	By using \eqref{eq_h_k} and assuming the users clustered in the same groups experience similar AoDs \cite{ASIL_Xiaoyi_DL_CE}, the channel matrix for group $g$ is given by:
	\begin{equation}\label{eq_H_g}
		{\bf H}_g 
		= 
		\big[
		{\bf h}_{g_1},\cdots,{\bf h}_{g_{K_g}}
		\big]^T
		= 
		{\bf Z}_g{\bf\Phi}_g
		\in
		\mathbb{C}^{K_g\times M},
	\end{equation}
	where 
	$g_k=k+\sum_{t=1}^{g-1}K_t$ is the UE index with $K\hspace{-0.5ex}=\sum_{g=1}^{G}K_g$,
	${\bf Z}_g = \big[
	{\bf z}_{g_1},\cdots,{\bf z}_{g_{K_g}}
	\big]^T\in\mathbb{C}^{K_g\times Q}$ 
	is the fast time-varying path gain matrix,
	${\bf \Phi}_g\in\mathbb{C}^{Q\times M}$ 
	is the slow time-varying phase response matrix.
	Afterwards, the concatenated full-size channel matrix is defined as 
	${\bf H} =
	\left[
	{\bf H}_1^T,
	\cdots,
	{\bf H}_G^T
	\right]^T
	\hspace{-0.5ex}\in\hspace{-0.25ex}\mathbb{C}^{K\times M}$.
	
	Then, $G$ blocks are designed for the RF beamformer as:
	\begin{equation}\label{eq_F}
		{\bf F}= \left[{\bf F}_1,\cdots,{\bf F}_G\right]\in \mathbb{C}^{M\times N_{RF}},
	\end{equation}
	where 
	${\bf F}_g \hspace{-0.25ex}\in \hspace{-0.25ex}\mathbb{C} ^{M\times N_{RF,g}}$ is the RF beamformer for group $g$
	with 
	$N_{RF}\hspace{-0.45ex}=\hspace{-0.55ex}\sum\nolimits_{g=1}^{G}\hspace{-0.25ex}N_{RF,g}$.
	By using \eqref{eq_H_g} and \eqref{eq_F}, the effective channel matrix seen from the BB-stage is obtained as:
	\begin{equation}\label{eq_H_eff}
		\tilde{\bf{H}}
		\hspace{-.25ex}=\hspace{-.25ex}
		{\bf H}
		{\bf F} 
		\hspace{-.25ex}=\hspace{-.5ex}
		\left[ \hspace{-1ex}{\begin{array}{*{20}{c}}
				{{{\bf{H}}_1}{{\bf{F}}_1}}&\hspace{-1ex}{{{\bf{H}}_1}{{\bf{F}}_2}}&\hspace{-1ex}{\cdots}&\hspace{-1ex}{{{\bf{H}}_1}{{\bf{F}}_G}}\\
				{{{\bf{H}}_2}{{\bf{F}}_1}}&\hspace{-1ex}{{{\bf{H}}_2}{{\bf{F}}_2}}&\hspace{-1ex}{\cdots}&\hspace{-1ex}{{{\bf{H}}_2}{{\bf{F}}_G}}\\
				{\vdots}&\hspace{-1ex}{\vdots}&\hspace{-1ex}{\ddots}&\hspace{-1ex}{\vdots}\\
				{{{\bf{H}}_G}{{\bf{F}}_1}}&\hspace{-1ex}{{{\bf{H}}_G}{{\bf{F}}_2}}&\hspace{-1ex}{\cdots}&\hspace{-1ex}{{{\bf{H}}_G}{{\bf{F}}_G}}\\
		\end{array}} \hspace{-1ex} \right]\hspace{-.5ex}\in\mathbb{C}^{K\times N_{RF}},
	\end{equation}
	where ${\bf H}_g{\bf F}_g={\bf Z}_g{\bf \Phi}_g{\bf F}_g\in\mathbb{C}^{K_g\times N_{RF,g}}$ is the effective channel matrix for group $g$ and ${\bf H}_t{\bf F}_g\hspace{-0.5ex}=\hspace{-0.5ex}{\bf Z}_t{\bf \Phi}_t{\bf F}_g\in\mathbb{C}^{K_t\times N_{RF,g}}$ is the effective interference channel matrix, $\forall t\neq g$.

	Hence, the RF beamformer design targets accomplishing the following two objectives: (i) maximizing the beamforming gain in the desired direction 
	(i.e., $\textrm{Span}\left({\bf F}_g\right) \subset  \textrm{Span}\left({\bf \Phi}_g\right)$),
	(ii) successfully suppress the interference among UE groups 
	(i.e., $\textrm{Span}\left({\bf F}_g\right) \subset  \cup_{t\ne g}\textrm{Null}\left({\bf \Phi}_t\right)$).
	As proven in \cite{ASIL_ABHP_Access}, both objectives are accomplished by building the RF beamformer ${\bf F}_g$ via the steering vector 
	${\bf e}\left(\gamma_x,\gamma_y\right)\hspace{-0.25ex}=\hspace{-0.25ex}\frac{1}{\sqrt{M}}{\bm{\phi}}^*\hspace{-0.25ex}\left( {{\gamma_x, \gamma_y}} \right)\hspace{-.25ex}\in\hspace{-0.25ex}\mathbb{C}^M$ with $\left(\gamma_x,\gamma_y\right)$ angle-pairs covering the AoD support of desired UE group and excluding the AoD supports of the other UE groups
	(please see \eqref{eq_phase_vector} for ${\bm{\phi}}\hspace{-0.25ex}\left( {{\gamma_x, \gamma_y}} \right)$).
	For covering the complete 3D elevation and azimuth angular space with minimum number of angle-pairs, $M$ orthogonal quantized angle-pairs are defined as
	${{\lambda^x_{u}} \hspace{-0.5ex}=\hspace{-0.5ex} -1 + \frac{2u-1}{{{M_{x}}}}}$ 
	for 
	$u = 1, \cdots,{M_{x}}$
	and 
	${{\lambda^y_{c}} = -1 + \frac{2c-1}{{{M_{y}}}}}$ 
	for
	$c = 1, \cdots,{M_{y}}$.
	Considering that $N_{RF,g}$ quantized angle-pairs covers the AoD support of  group $g$ \cite[eq. (13)]{ASIL_ABHP_Access}, we build the RF beamformer for UE group $g$ as follows:
	\begin{equation}\label{eq_F_g}
		{\bf F}_g
		\hspace{-0.5ex}=\hspace{-0.5ex}
		\big[\hspace{-0.25ex}{\bf e}\big(\hspace{-.25ex}\lambda_{u_1}^x,\hspace{-0.25ex}\lambda_{c_1}^y\hspace{-0.25ex}\big),\hspace{-0.25ex}\cdots\hspace{-0.25ex},
		\hspace{-.25ex}{\bf e}\big(\hspace{-.25ex}\lambda_{u_{N_{RF,g}}}^x\hspace{-.5ex},\hspace{-.25ex}\lambda_{c_{N_{RF,g}}}^y\hspace{-.25ex}\big)\hspace{-.25ex}\big]\hspace{-.5ex}\in\hspace{-.5ex}\mathbb{C}^{M\hspace{-.25ex}\times N_{RF,g}}\hspace{-.25ex}.
	\end{equation}
	Finally, the complete RF beamformer ${\bf F}$ satisfying the CM constraint (i.e., $C_3$ given in \eqref{eq_OPT_1}) is derived by  substituting \eqref{eq_F_g} into \eqref{eq_F}.
	It is worthwhile to mention that the analog RF beamformer is a unitary matrix (i.e., ${\bf F}^H {\bf F} = {\bf I}_{N_{RF}}$).
	
	\subsection{Digital BB Precoder}\label{SSEC_BB}
	We aim to further mitigate the residual inter UE interference at the digital BB precoder.
	Thus, the regularized zero-forcing (RZF) technique is applied via joint group processing \cite{ASIL_ABHP_Access}. 
	By utilizing the reduced-size effective channel matrix $\tilde{\bf{H}}$ defined in \eqref{eq_H_eff}, the digital BB precoder is constructed as \cite{Mass_MIMO_Precoding_Survey}:
	\begin{equation}\label{eq_BB}
		{\bf B}
		= \left[
		\tilde{\bf{H}}^H
		\tilde{\bf{H}}
		+ 
		K
		\frac{\sigma_n^2}{P_T}
		{\bf I}_{N_{RF}}
		\right]^{-1}
		\tilde{\bf{H}}^H
		\in
		\mathbb{C}^{N_{RF}\times K}.
	\end{equation}
	
	\vspace{1.5ex}
	\section{A Low-Complexity Deep Learning based \\Power Allocation} \label{SEC_DL_PA}
	
	After developing the analog RF beamformer $\bf F$ and the digital BB precoder $\bf B$, the capacity maximization optimization problem given in \eqref{eq_OPT_1} is reformulated as follows:
	\begin{equation}\label{eq_OPT_2}
		\begin{aligned}
			\max_{
				{\bf P}
			} ~& 
			\sum_{k=1}^{K}
			\log_2
			\left(
			1
			+
			\frac
			{p_k\left|{\bf h}_k^T{\bf Fb}_k\right|^2}
			{\sum_{t\ne k}^{K}p_t\left|{\bf h}_k^T{\bf Fb}_t\right|^2 
				+
				\sigma_n^2}
			\right)\\
			\textrm{s.t.}~~~
			&
			C_1: \mathbb{E}\big\{\hspace{-0.5ex}\left\|{\bf{s}}\right\|^2_2\hspace{-0.5ex}\big\} = \sum_{k = 1}^K {{p_{k}}{\bf{b}}_{k}^H{{\bf{F}}^H}{\bf{F}}{{\bf{b}}_{k}}} \le {P_T},\\
			&
			C_2: p_{k}\ge 0, \forall k,
		\end{aligned}
	\end{equation}
	However, it is still a non-convex optimization problem due to the optimization variables as ${\bf P}=\textrm{diag}\left(\sqrt{p}_1,\cdots,\sqrt{p}_K\right)$ interchangeably located in the numerator and denominator \cite{Emil_PA_NonConvex}.
	Thus, the traditional optimization algorithms may not be utilized to solve the PA problem. 
	
	Recently, a particle swarm optimization\footnote{As a nature-inspired AI technique, the particle swarm optimization (PSO) employs multiple search agents (i.e., particles), which communicate and move through iterations with the goal of finding the globally optimal solution\cite{PSO_book}.} based power allocation (PSO-PA)\footnote{The details of PSO-PA algorithm {are} available in \cite[Algorithm 1]{ASIL_PSO_PA_WCNC}.} technique for finding the optimal allocated powers is proposed in \cite{ASIL_PSO_PA_WCNC}. In comparison to the computationally expensive exhaustive search, it is numerically shown that the global optimal solution is achieved via PSO-PA.
	However, as the number of UEs increases (i.e., higher dimensional optimization space), PSO-PA requires more iterations  and longer runtime. Thus, the enhanced computational complexity might make PSO-PA impractical for the real-time online applications of MU-{m}MIMO systems.
	
	For achieving a near-optimal sum-rate performance while keeping {a} reasonable runtime, we propose a low-complexity deep learning based power allocation (DL-PA) algorithm. Here, we have two phases as demonstrated in Fig. \ref{fig2_Learning_Phases}:
	(i) Phase 1 applies the offline supervised learning via the optimal allocated power values calculated by PSO-PA,
	(ii) Phase 2 runs the trained DL-PA algorithm for predicting the allocated powers in the real-time online applications.
	
	Hence, the reminder of this section introduces the DNN architecture, loss functions, dataset generation and training process for the proposed low-complexity DL-PA algorithm.
	
	\subsection{Deep Neural Network Architecture}
	
\begin{figure}
		\centering
		\includegraphics[width = \columnwidth]{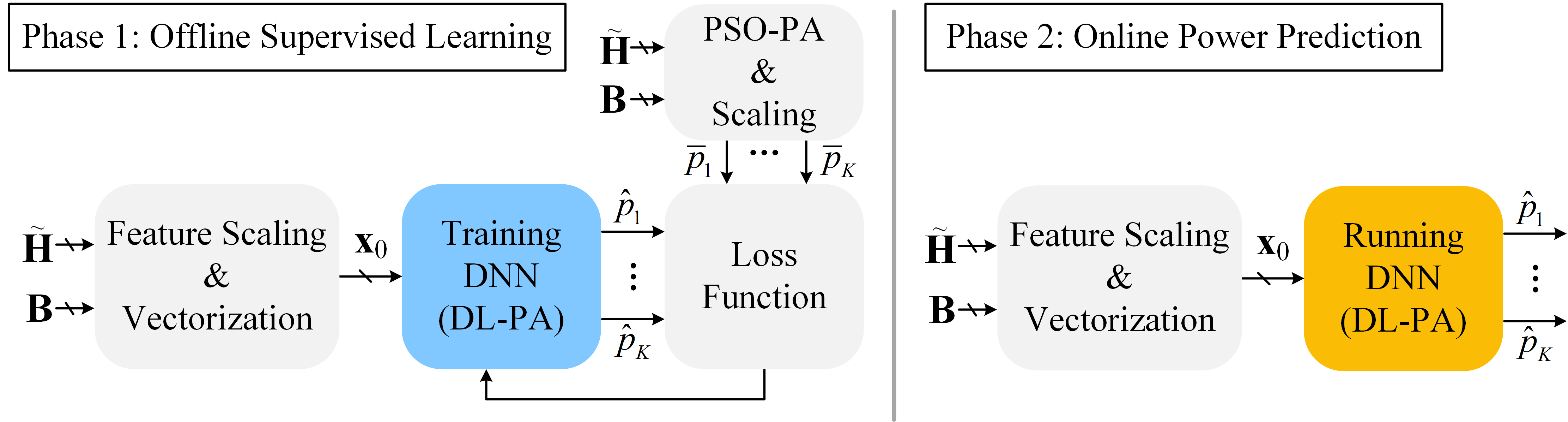}
		\caption{Block diagram of offline supervised learning (Phase 1) and online power prediction (Phase 2) in the DL-PA algorithm.}
		\label{fig2_Learning_Phases}
	\end{figure}

\begin{figure}
		\centering
		\includegraphics[width = \columnwidth]{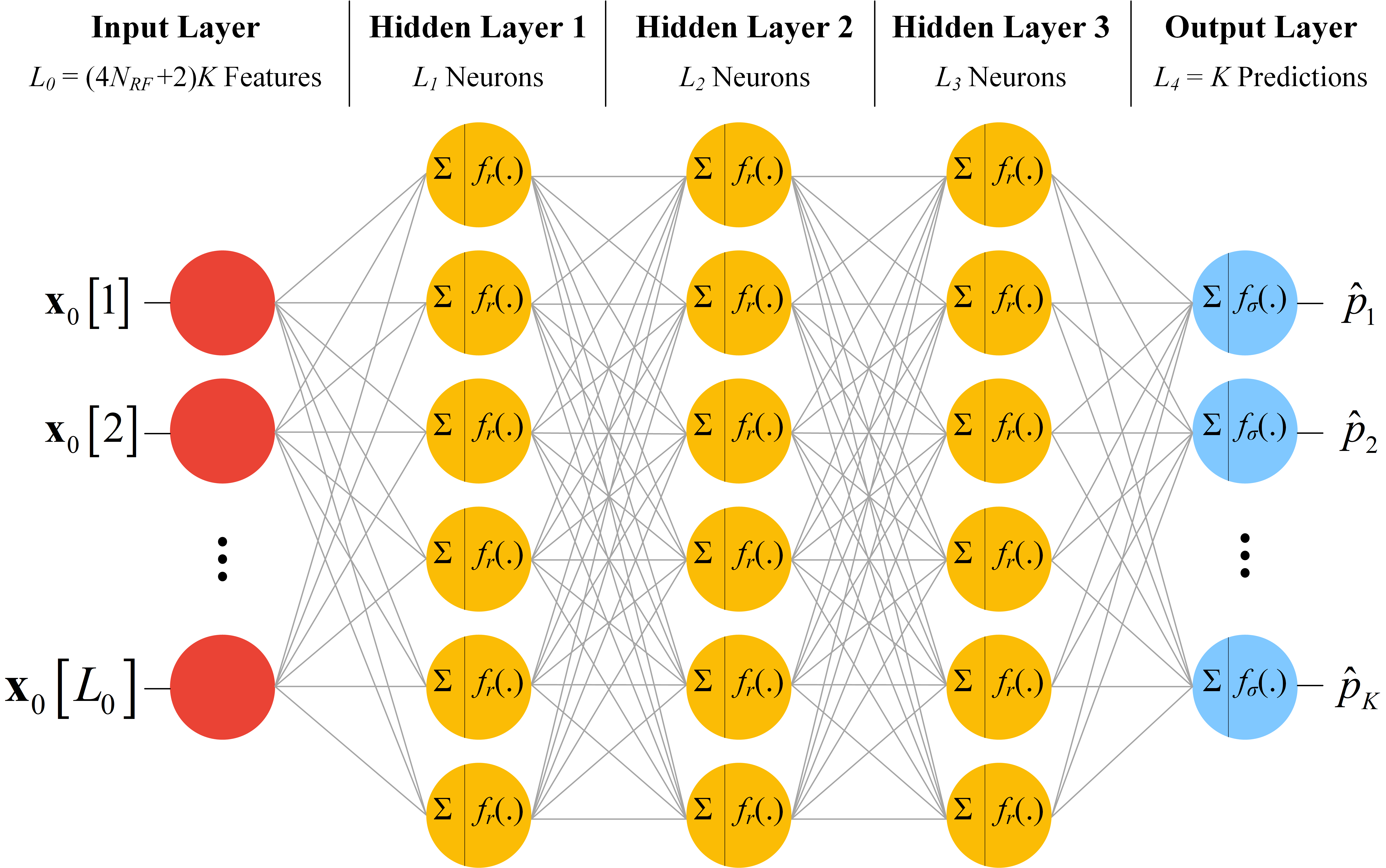}
		\caption{Deep neural network architecture for DL-PA algorithm.}
		\label{fig3_DNN}
	\end{figure}

	We model a fully-connected deep neural network (DNN) architecture with three hidden layers as illustrated in Fig. \ref{fig3_DNN}, which aims to predict the optimal allocated powers for $K$ downlink UEs. There are $L_i$ neurons present at the $i^{th}$ hidden layer with $i=1,2,3$. 
	On the other hand, as shown in Fig. \ref{fig2_Learning_Phases}, the effective channel matrix 
	$\tilde{\bf H}={\bf HF}=\big[\tilde{\bf h}_1^T,\cdots,\tilde{\bf h}_K^T\big]\in\mathbb{C}^{K \times N_{RF}}$ given in  \eqref{eq_H_eff}
	and the digital BB precoder 
	${\bf B}=\big[{\bf b}_1,\cdots,{\bf b}_K\big]\in\mathbb{C}^{N_{RF} \times K}$ 
	given in \eqref{eq_BB}
	are employed as inputs in the proposed DL-PA algorithm.
	The input feature scaling and vectorization operations are applied to $\tilde{\bf H}$ and ${\bf B}$. 
	Then, the input layer feature vector is obtained as:
	\begin{equation} \label{eq_x_input}
		{\bf x}_0 = 
		\left[
		\begin{array}{*{20}{l}}
			\alpha_{1}{\bf x}_{\tilde{\bf h}_1}\\
			\hspace{3ex}\vdots\\
			\alpha_{1}{\bf x}_{\tilde{\bf h}_K}\\
			\alpha_{2}{\bf x}_{{\bf b}_1}\\
			\hspace{3ex}\vdots\\
			\alpha_{2}{\bf x}_{{\bf b}_K}\\
			\alpha_{3}{\bf {x}}_{\textrm{BB}}\\
			\alpha_{4}{\bf {x}}_{\textrm{BB,inv}}
		\end{array}
		\hspace{-1ex}\right]
		\in
		\mathbb{R}^{L_0},
	\end{equation}
	where 
	$L_0=\left(4N_{RF}+2\right)K$ is the input feature size,
	$
	{\bf x}_{\tilde{\bf h}_k}
	=
	\big[
	\operatorname{Re}\big(\tilde{\bf {h}}_{k}^T\big),
	\operatorname{Im}\big(\tilde{\bf {h}}_{k}^T\big)
	\big]^T
	\in
	\mathbb{R}^{2N_{RF}}
	$,
	$
	{\bf x}_{{\bf b}_k}
	=
	\big[
	\operatorname{Re}\big({\bf {b}}_{k}^T\big),
	\operatorname{Im}\big({\bf {b}}_{k}^T\big)
	\big]^T
	\in
	\mathbb{R}^{2N_{RF}}
	$,
	${\bf {x}}_{\textrm{BB}}^T
	=
	\big[
	{{\bf b}_1^H{\bf b}_1},
	\cdots,
	{{\bf b}_K^H{\bf b}_K}
	\big]^T
	\in
	\mathbb{R}^{K}
	$
	and
	${\bf {x}}_{\textrm{BB,inv}}^T
	=
	\big[
	\frac{1}{{\bf b}_1^H{\bf b}_1},
	\cdots,
	\frac{1}{{\bf b}_K^H{\bf b}_K}
	\big]^T
	\in
	\mathbb{R}^{K}
	$
	are respectively the non-scaled input feature vectors for the effective channel, BB precoder, the gain of each BB precoder vector and its inverse.
	By implementing the maximum absolute scaling \cite{MaxAbsScalar}, the corresponding scaling coefficients are calculated as:
	\begin{equation}
		\begin{aligned}
			\alpha_{1}
			&=
			\max
			\left(
			\big|{\bf x}_{\tilde{\bf h}_1}^T\big|,
			\cdots,
			\big|{\bf x}_{\tilde{\bf h}_K}^T\big|
			\right)^{-1}\\
			\alpha_{2}
			&=
			\max
			\left(
			\big|{\bf x}_{{\bf b}_1}^T\big|,
			\cdots,
			\big|{\bf x}_{{\bf b}_K}^T\big|
			\right)^{-1}\\
			\alpha_{3}
			&=
			\max
			\left(
			{{\bf b}_1^H{\bf b}_1},
			\cdots,
			{{\bf b}_K^H{\bf b}_K}
			\right)^{-1}\\
			\alpha_{4}
			&=
			\min
			\left(
			{{\bf b}_1^H{\bf b}_1},
			\cdots,
			{{\bf b}_K^H{\bf b}_K}
			\right).
		\end{aligned}
	\end{equation}
	Hence, each element of the input feature vector is scaled between $-1$ and $1$ (i.e., 
	${\bf x}_0\in\left[-1,1\right]$) 
	by the maximum absolute scaling technique. It prevents the domination of large valued features on the small valued features \cite{MaxAbsScalar}. 
	
	In the offline supervised learning process (i.e., Phase 1), the optimal allocated powers are calculated as the output labels via the computationally expensive PSO-PA algorithm. Similar to the input features, we also apply the maximum absolute scaling to the optimal allocated powers as follows:
	\begin{equation}\label{eq_pk_label}
		\bar{p}_k=\frac{p_{k}^{\textrm{opt}}}
		{\max\left({p_{1}^{\textrm{opt}}},\cdots,{p_{K}^{\textrm{opt}}}\right)} \in\left[0, 1\right].
	\end{equation}
	
	For the non-linear operations, we utilize the rectified linear unit (ReLU) as the activation function at the hidden layers (i.e., 
	$f_r\left(x\right)=\max\left(0,x\right)$ \cite{lecun2015deep}).
	Therefore, by using the input feature vector 
	${\bf x}_0$ 
	given in \eqref{eq_x_input},
	the output of $i^{th}$ hidden layer is calculated as 
	${{\bf x}_{i} = f_r\left({\bf W}_{i-1}{\bf x}_{i-1} + {\bf b}_{i-1}\right)\in\mathbb{R}^{L_{i}}}$,
	where 
	${\bf W}_{i-1}\in\mathbb{R}^{L_{i}\times L_{i-1}}$ 
	and 
	${\bf b}_{i-1}\in\mathbb{R}^{L_{i}}$
	are the weight matrix and bias vector, respectively.
	In order to fit the output layer predictions between $0$ and $1$ as in the output labels expressed in \eqref{eq_pk_label}, we employ the sigmoid function at the output layer (i.e., $f_{\sigma}\left(x\right)=\frac{1}{1+e^{-x}}$ \cite{lecun2015deep}).
	Thus, the predicted power values for $K$ downlink UEs via the DNN architecture are written as:
	\begin{equation}\label{eq_predicted_power}
		\begin{aligned}
			\left[
			\right.
			\hspace{-0.75ex}&\hspace{0.75ex}
			\left.\hspace{-0.35ex}
			\hat{p}_1,
			\hat{p}_2,
			\cdots,
			\hat{p}_K
			\right]\\
			&
			=\hspace{-0.5ex}
			f_{\sigma}
			\hspace{-0.5ex}
			\left(
			{\bf W}_{\hspace{-0.35ex}3}
			{\bf x}_3
			\hspace{-0.35ex}+\hspace{-0.35ex}
			{\bf b}_3\right)\\
			&
			=\hspace{-0.5ex}
			f_{\sigma}
			\hspace{-0.5ex}
			\left(
			{\bf W}_{\hspace{-0.35ex}3}
			f_{r}
			\hspace{-0.5ex}
			\left(
			{\bf W}_{\hspace{-0.35ex}2}
			f_{r}
			\hspace{-0.5ex}\left(
			{\bf W}_{\hspace{-0.35ex}1}
			f_{r}
			\hspace{-0.5ex}\left(
			{\bf W}_{\hspace{-0.35ex}0}
			{\bf x}_0
			\hspace{-0.5ex}+\hspace{-0.5ex}
			{\bf b}_0\right)
			\hspace{-0.5ex}+\hspace{-0.5ex}
			{\bf b}_1\right)
			\hspace{-0.5ex}+\hspace{-0.5ex}
			{\bf b}_2\right)
			\hspace{-0.5ex}+\hspace{-0.5ex}
			{\bf b}_3\right)\hspace{-0.5ex}.
		\end{aligned}
	\end{equation}
	By using \eqref{eq_BB} and \eqref{eq_predicted_power}, we finally derive the multi-user PA matrix satisfying the transmit power constraint of $P_T$ as:
	\begin{equation}
		{\bf P} = 
		\sqrt{
			\frac{P_T}{\sum\nolimits_{k=1}^{K}\hat{p}_k{\bf b}_k^H{\bf b}_k}
		}
		\textrm{diag}\left(\sqrt{\hat{p}_1},\sqrt{\hat{p}_2},\cdots,\sqrt{\hat{p}_K}\right).
	\end{equation}

	\subsection{Loss Functions}
	We here consider two loss functions by using the predicted and optimal power values: (i) mean square error (MSE), (ii) mean absolute error {(MAE)}. When there are $S$ network realizations in the dataset, the MSE loss {function} is given by:
	\begin{equation}\label{eq_MSE}
		\begin{aligned}
			\mathcal{L}_{\textrm{MSE}}
			&=
			\frac{1}{SK}\sum_{i=1}^{S}\sum_{{k=1}}^{K}
			\left(\bar{p}_{k,i} - \hat{p}_{k,i}\right)^2.
		\end{aligned}
	\end{equation}	
	Similarly, the MAE loss function is written as:
	\begin{equation}\label{eq_MAE}
		\begin{aligned}
			\mathcal{L}_{\textrm{MAE}}
			&=
			\frac{1}{SK}\sum_{i=1}^{S}\sum_{{k=1}}^{K}
			\left|\bar{p}_{k,i} - \hat{p}_{k,i}\right|.
		\end{aligned}
	\end{equation}
	By back-propagating the gradients of loss function  from the output layer to the input layer, 
	the weight matrices ${\bf W}_i$ and bias vectors ${\bf b}_i$ are updated for reducing the loss and closely predicting the optimal allocated power values. Hence, we ultimately optimize the sum-rate capacity of MU-{m}MIMO systems as expressed in \eqref{eq_OPT_2}.

	\subsection{Dataset Generation \& Training Process}
	
	We generate a dataset with $S=100.000=10^5$ network realizations for the offline supervised learning process (i.e., Phase 1) illustrated in Fig. \ref{fig2_Learning_Phases}. In each realization, the channel vector expressed in \eqref{eq_h_k} is generated for each UE by randomly varying the path gains, AoD parameters and UE location with respect to the BS.
	The corresponding optimal allocated powers are calculated via the PSO-PA algorithm \cite[Algorithm 1]{ASIL_PSO_PA_WCNC} and stored in the dataset.
	For the offline learning process, we always consider $80\%$-$20\%$ split of the total available dataset among the training and validation.
	After completing the offline learning process (i.e., Phase 1), the online power allocation (i.e., Phase 2) is tested with a purely new test dataset.
		The DNN architecture for the proposed DL-PA algorithm is implemented using the open-source deep learning libraries in TensorFlow \cite{TensorFlow}. 
	
	\section{Illustrative Results}\label{SEC_RESULTS}

	This section presents sum-rate and runtime results for evaluating the proposed AB-HP with deep learning based power allocation (DL-PA) in the MU-{m}MIMO systems.
	The simulation parameters according to the 3D microcell scenario are summarized in Table \ref{table1_sim}\footnote{When a square URA having $256$ antennas is utilized to serve $G=1$ UE group, AB-HP reduces the number of RF chains from $256$ to $12$ according to the given simulation setup. It means $95.3\%$ reduction in the number of RF chains and channel estimation overhead compared to the conventional FDP.}.
	Furthermore, the hyper-parameters for the DNN architecture are outlined in Table \ref{table2_DL_PA}.
	\begin{table}[t!]
		\caption{Simulation parameters.}
		\vspace{-1ex}
		\label{table1_sim}
		\centering
		\renewcommand{\arraystretch}{1.25}
		\begin{tabular}{|l|l|}
			\hline
			{Number of antennas \cite{Report_5G_UMi_UMa_Rel16}} & $M=16\times 16 = 256$      \\ \hline
			{BS transmit power  \cite{Report_5G_UMi_UMa_Rel16}} &  $P_T=20$ dBm \\ \hline
			{Cell radius  \cite{Report_5G_UMi_UMa_Rel16}} &  100m\\ \hline
			{BS height \cite{Report_5G_UMi_UMa_Rel16} \big|\big. UE height \cite{Report_5G_UMi_UMa_Rel16}} &  10m \big|\big. 1.5m-2.5m \\ \hline
			{UE-BS horizontal distance} & 10m -- 90m\\ \hline
			{UE groups} & {$G=1$ or $G=2$}\\ \hline
			{UE per group} & $K_g=\frac{K}{G}$      \\ \hline
			{Mean EAoD \big|\big. Mean AAoD} & {$\theta_g \hspace{-0.5ex}=\hspace{-0.25ex} 60^\circ$ \hspace{-0.5ex}\big|\big. \hspace{-0.5ex}$\psi_g\hspace{-0.5ex}=\hspace{-0.5ex}21^\circ \hspace{-1ex}+\hspace{-0.5ex} 180^\circ\hspace{-0.5ex}\left(g\hspace{-0.25ex}-\hspace{-0.25ex}1\right)$ }  \\ \hline
			EAoD spread \big|\big. AAoD spread & $\delta_g^\theta\hspace{-0.5ex}=\hspace{-0.5ex}15^\circ$ \hspace{-0.25ex}\big|\big. $\delta_g^\psi=11^\circ$\hspace{10.6ex} \\ \hline
			{Path loss exponent} \cite{Report_5G_Macro_PL_Rel_16}& $\eta=3.76$      \\ \hline
			{Noise PSD} \cite{Report_5G_Macro_PL_Rel_16}& $-174$ dBm/Hz      \\ \hline
			{Channel bandwidth} \cite{Report_5G_Macro_PL_Rel_16}& $10$ kHz      \\ \hline
			{\# of paths\cite{Report_5G_UMi_UMa_Rel16}} & $Q=20$      \\ \hline
			Antenna spacing (in wavelength) & ${d\hspace{-0.25ex}=0.5}$\\ \hline
		\end{tabular}
	\end{table}
	
	\begin{table}[t!]
		\caption{DNN hyper-parameters.}
		\vspace{-1ex}
		\label{table2_DL_PA}
		\centering
		\renewcommand{\arraystretch}{1.25}
		\begin{tabular}{|l|l|}
			\hline
			$1^{th}$ hidden layer size & $L_1=1024$ \\ \hline
			$2^{th}$ hidden layer size & $L_2=512$ \\ \hline
			$3^{th}$ hidden layer size & $L_3=256$ \\ \hline
			{Dataset size} & $S=100.000$      \\ \hline
			{Test dataset size} & $1.000$      \\ \hline
			{Epoch size \big|\big. Batch size} & $25$ \big|\big. $32$      \\ \hline
			{Learning rate} & $0.001$      \\ \hline
			{Optimizer} & ADAM \cite{TensorFlow}      \\ \hline
		\end{tabular}
	\vspace{-2ex}
	\end{table}
	
\begin{figure}[!t]
		\begin{subfigure}{\columnwidth}
			\centering
			\includegraphics[width = \textwidth]{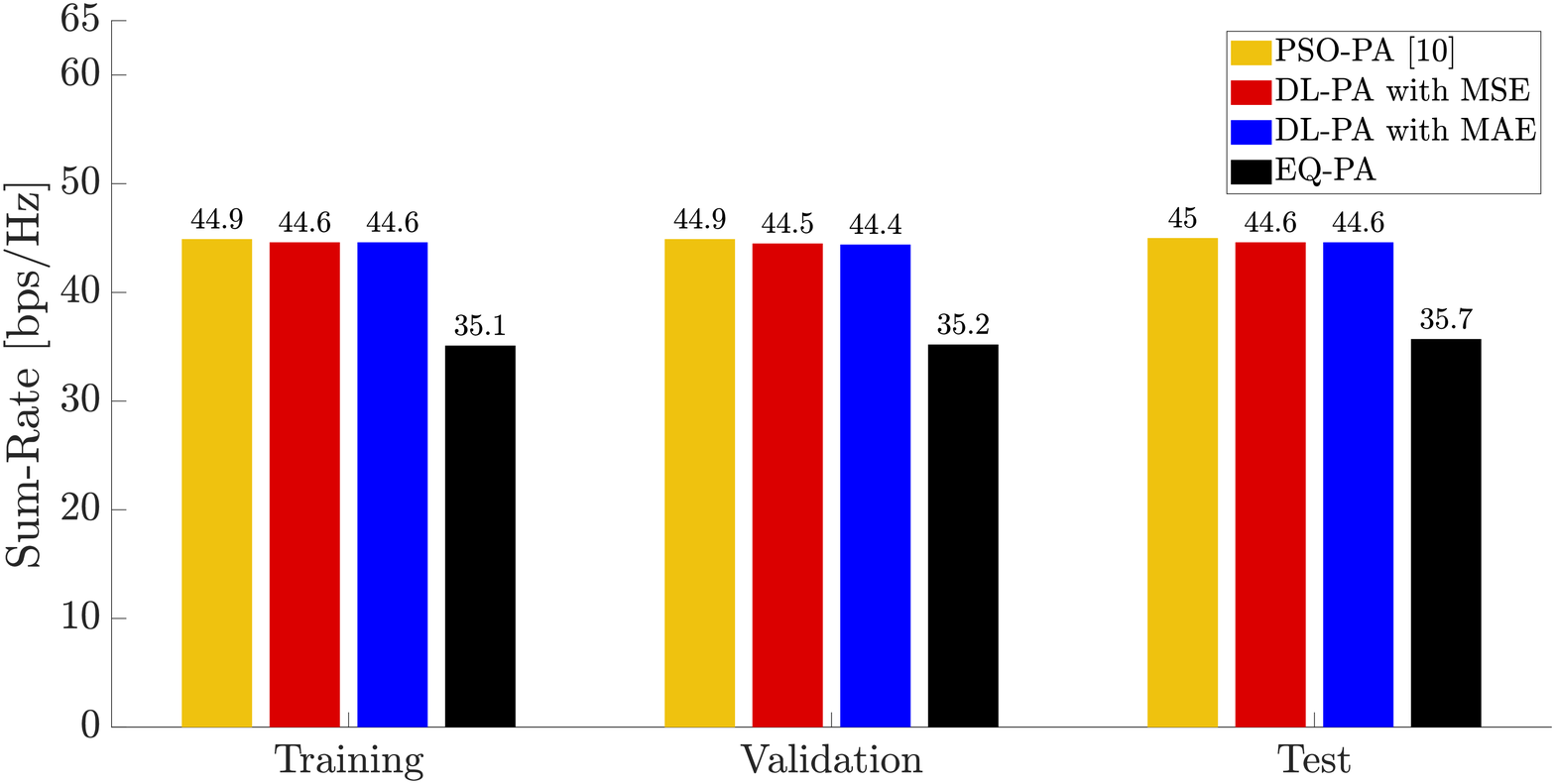}
			\vspace{-3ex}
			\caption{$K=3$}
		\end{subfigure}
		\begin{subfigure}{\columnwidth}
			\centering
			\includegraphics[width = \textwidth]{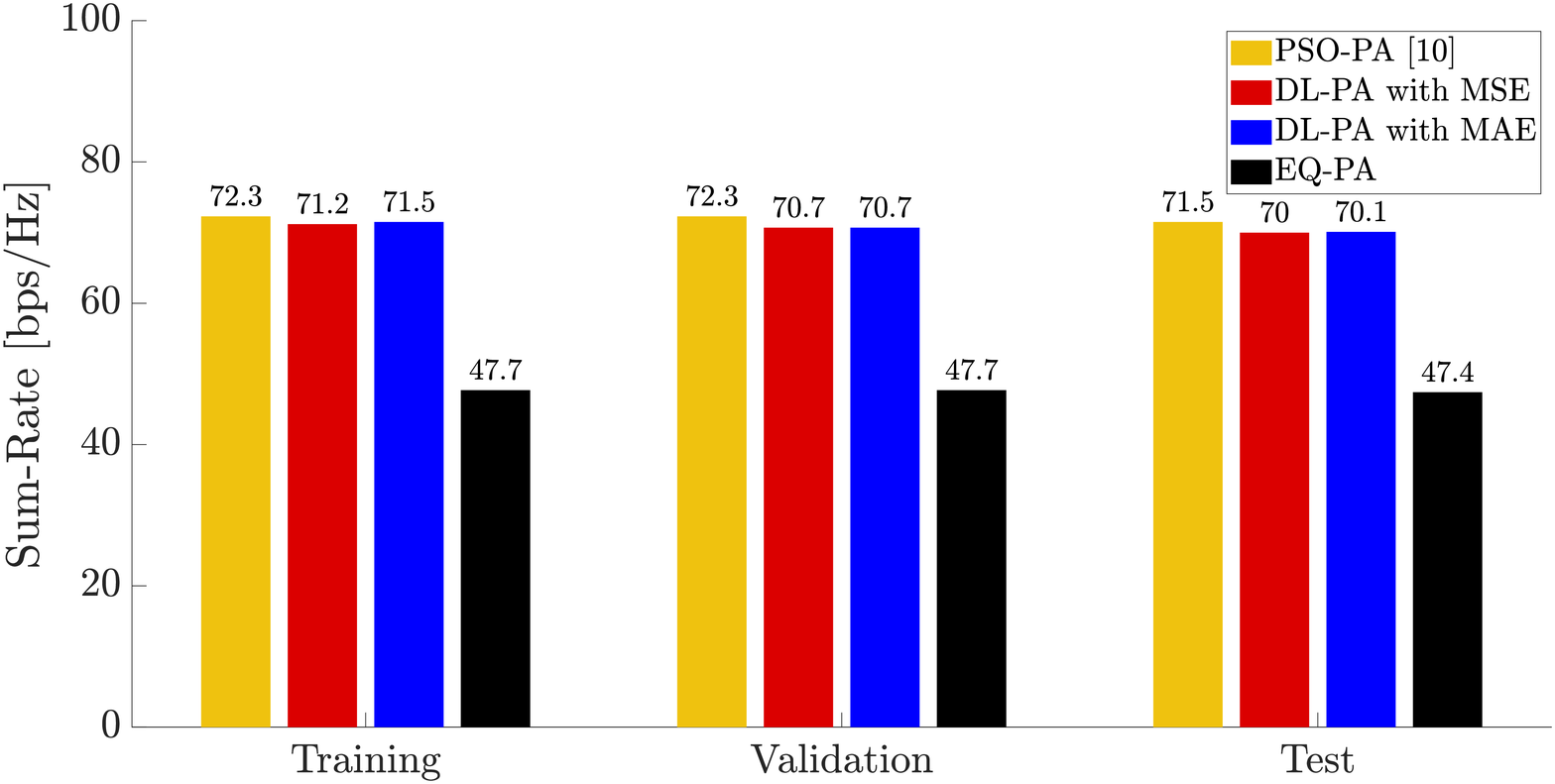}
			\vspace{-3ex}
			\caption{$K=6$}
		\end{subfigure}
		\vspace{-1ex}
		\caption{Sum-rate performance evaluation on training, validation and test dataset ($G=1$ group).}
		\label{fig4}
		\vspace{-2ex}
	\end{figure}

	Fig. \ref{fig4} plots the sum-rate of the proposed DL-PA with MSE and MAE loss functions defined in \eqref{eq_MSE} and \eqref{eq_MAE}, respectively. Here, we provide the performance evaluation on training, validation and test dataset for $K=3$ and $K=6$ UEs in $G=1$ group. 
	As a benchmark, DL-PA is compared with PSO-PA \cite{ASIL_PSO_PA_WCNC} and equal PA (EQ-PA).
	Numerical results {reveal} that the proposed $\textrm{DL-PA}$ closely approaches PSO-PA in all training, validation and test. 
	For instance, when there are $K=3$ UEs, DL-PA provides $44.6$ bps/Hz sum-rate capacity on test data and achieves $99.1\%$ of the optimal sum-rate capacity achieved by PSO-PA as $45$ bps/Hz.
	Additionally, the capacity is improved by approximately $25\%$ with respect to EQ-PA 
	(i.e., from $35.7$ bps to $44.6$ bps/Hz).
	Moreover, when there are {a} larger number of UEs as $K=6$, the sum-rate improvement compared to EQ-PA increases $48.8\%$ on the test data (i.e., from $47.4$ bps to $70.1$ bps/Hz).
	However, as the number of UEs increases, the optimization space enlarges and we observe a slight decay {in} the test data performance. To illustrate, for $K=6$ UEs, DL-PA with MAE accomplishes $98.9\%$ of the optimal sum-rate performance on training data, which marginally drops to $98.1\%$ on the test data. 
	
\begin{figure}[!t]
		\centering
		\includegraphics[width = \columnwidth]{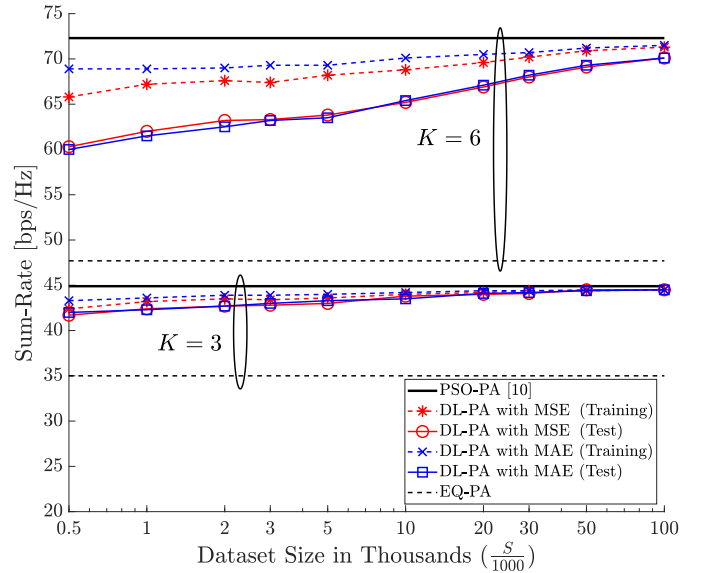}
		\vspace{-3ex}
		\caption{Sum-rate performance versus dataset size \hspace{-0.25ex}($G\hspace{-0.5ex}=\hspace{-0.5ex}1$ group).}
		\label{fig5}
		\vspace{-2ex}
	\end{figure}

	In Fig. \ref{fig5}, the sum-rate performance is demonstrated versus the dataset size $S$, where there are either $K=3$ or $K=6$ UEs in $G=1$ group and dataset size varies between $500$ and $100.000$. It is seen that as the dataset size increases the gap between PSO-PA and DL-PA vanishes. As expected, the larger dataset size makes DL-PA learn better the optimal allocated powers, especially on the unseen test dataset.
	
\begin{figure}[!t]
		\begin{subfigure}{0.48\columnwidth}
			\centering
			\includegraphics[width = \textwidth]{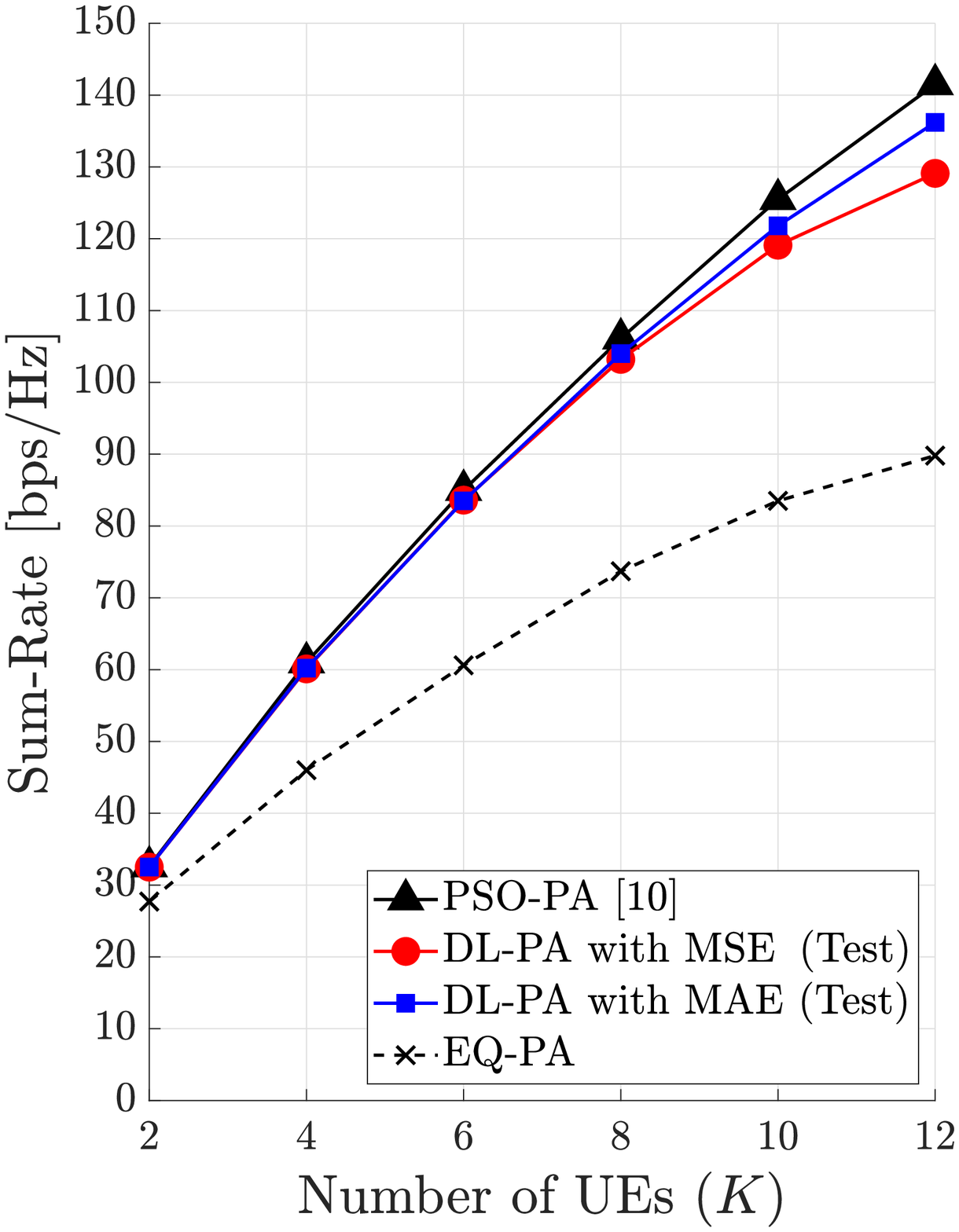}
			\caption{Sum-rate}
		\end{subfigure}
		\hfill
		\begin{subfigure}{0.48\columnwidth}
			\centering
			\includegraphics[width = \textwidth]{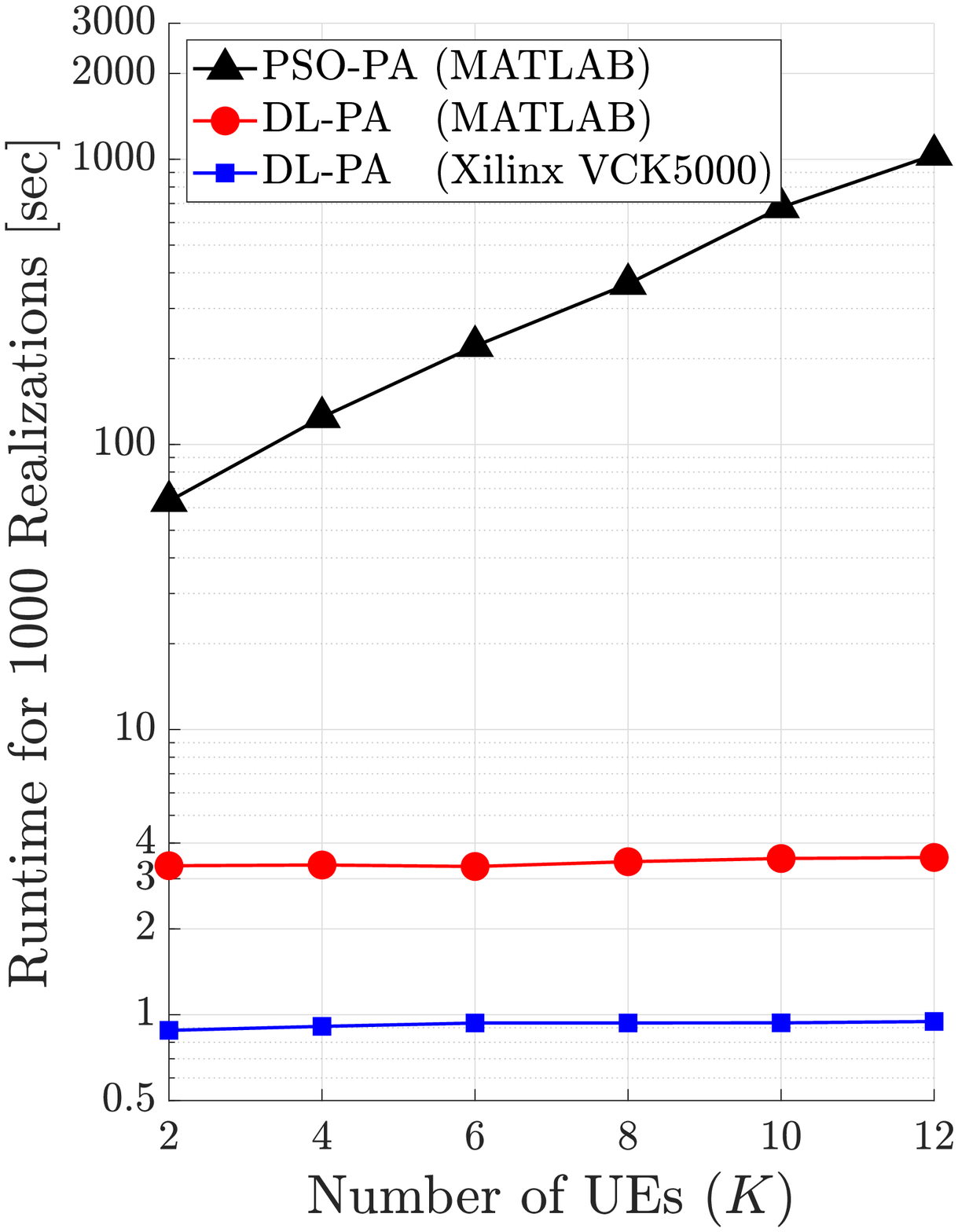}
			\caption{Runtime}
		\end{subfigure}
		\vspace{-1ex}
		\caption{Sum-rate and runtime performance ($G=2$ groups).}
		\vspace{-2ex}
		\label{fig6}
	\end{figure}

	Fig. \ref{fig6} displays both sum-rate and runtime results versus the number of UEs, which are equally clustered in $G=2$ groups (i.e., $K_g=\frac{K}{2}$).
	As seen from Fig. \ref{fig6}(a), DL-PA with MAE outperforms its MSE counterpart as the number of UEs increases, although their performance difference is not distinguishable for {a} smaller number of UEs.
	On the other hand, the relative sum-rate performance of DL-PA with MAE compared to the optimal PSO-PA algorithm varies between $99.7\%$ and $96.5\%$  as shown in Table \ref{table3_Relative_Performance}.
	Moreover, the runtime comparison between PSO-PA and DL-PA is demonstrated for 1000 network realizations in Fig. \ref{fig6}(b). It is worthwhile to note that the offline trained DNN architecture for DL-PA algorithm is run on both MATLAB\footnote{For the MATLAB runtime results, we implement both PSO-PA and DL-PA via a PC with Intel Core(TM) i7-4770 CPU @ 3.4 GHz and 32 GB RAM.} and Xilinx VCK5000 development card for AI inference \cite{Xilinx_AI_Card}.
	We observe that the proposed DL-PA strikingly outperforms the computational complex PSO-PA algorithm by significantly reducing the runtime. To illustrate, when there are $K=12$ UEs, PSO-PA requires $1036.6$ sec, whereas only $0.9$ sec runtime is enough to run DL-PA on Xilinx VCK5000.
	Also, the runtime for DL-PA remains almost constant across all UE scenarios because the hidden layers {have} the same architecture for various UE cases (e.g., approximately $3.5$ sec on MATLAB and $0.9$ sec on Xilinx VCK5000). 
		{Thus, the runtime per realization is below 1 msec on Xilinx VCK5000.}
	However, when there are more UEs, the runtime for PSO-PA exponentially increases due to the larger optimization space, where PSO-PA requires more iterations with the aim of finding the global optimal sum-rate.
		As presented in Table \ref{table3_Relative_Performance}, the relative runtime of DL-PA with MAE is reduced by $98.6\%$ for $K=2$ ($99.9\%$ for $K=12$) in comparison to the computationally expensive PSO-PA.

	\section{Conclusions}\label{SEC_CONC}
	In this work, a novel deep learning based power allocation (DL-PA) and hybrid precoding technique has been proposed for maximizing sum-rate capacity in the MU-{m}MIMO systems. First, the angular-based hybrid precoding (AB-HP) scheme has been expressed for the downlink transmission to reduce the number of RF chains and lower the channel estimation overhead. Then, we have proposed the low-complexity DL-PA algorithm for predicting the optimal allocated power resources among the downlink UEs.
		The promising numerical results show that the proposed DL-PA closely  approaches the optimal sum-rate capacity achieved by PSO-PA.
	On the other hand, DL-PA greatly reduces the runtime by $98.6\%$-$99.9\%$. 
	It makes the implementation of DL-PA feasible for the real-time online applications in MU-{m}MIMO systems.
	
	\begin{table}[t!]
		\caption{Relative performance of DL-PA with \hspace{-0.25ex}MAE \hspace{-0.25ex}($G\hspace{-0.5ex}=\hspace{-0.5ex}2$).}
		\vspace{-1ex}
		\label{table3_Relative_Performance}
		\centering
		\renewcommand{\arraystretch}{1.25}
		\begin{tabular}{|l|c|c|c|c|c|c|}
			\hline
			& $K\hspace{-0.5ex}=\hspace{-0.5ex}2$
			& $K\hspace{-0.5ex}=\hspace{-0.5ex}4$
			& $K\hspace{-0.5ex}=\hspace{-0.5ex}6$
			& $K\hspace{-0.5ex}=\hspace{-0.5ex}8$
			& $K\hspace{-0.5ex}=\hspace{-0.5ex}10$
			& $K\hspace{-0.5ex}=\hspace{-0.5ex}12$ \\ \hline
			\hspace{-0.5ex}Sum-Rate\hspace{-1ex} &  $99.7\%$   &  $98.7\%$   & $98.3\%$  &  $98.0\%$ &  $97.1\%$ &  $96.5\%$\\ \hline
			\hspace{-0.5ex}Runtime\hspace{-1ex}  &  $1.39\%$   &  $0.73\%$   & $0.42\%$ & $0.26\%$ &  $0.14\%$ &  $0.09\%$\\ \hline
		\end{tabular}
		\vspace{-2.5ex}
	\end{table}
	
	\vspace{-0.8ex}
	\ifCLASSOPTIONcaptionsoff
	\newpage
	\fi
	\bibliographystyle{IEEEtran}
	\bibliography{bibAsil_2110}
	\balance

\end{document}